\documentclass[12pt]{article}
\usepackage{graphicx}

\usepackage{epsfig,amsmath,amssymb,verbatim,mathrsfs,subfigure}

\def\u1x{{U(1)_X}}
\def\beq{\begin{equation}}
\def\eeq{\end{equation}}
\def\bea{\begin{eqnarray}}
\def\eea{\end{eqnarray}}
\def\bmat{\begin{pmatrix}}
\def\emat{\end{pmatrix}}
\def\tev{\,{\rm TeV}}

\def\gev{\,{\rm GeV}}

\def\mz{{m_Z}}

\def\mh{{m_h}}
\def\mH{{m_H}}

\def\Chi{{\cal X}}
\def\psione{{\Psi_1}}
\def\psitwo{{\Psi_2}}

\def\mf{m_f}
\def\mw{m_W}
\def\mpsi{M_\psi}

\def\cL{{c_\psi}_L}
\def\sL{{s_\psi}_L}
\def\cR{{c_\psi}_R}
\def\sR{{s_\psi}_R}

\def\brinv{$BR_{\rm inv}\ $}

\newcommand{ \slashchar }[1]{\setbox0=\hbox{$#1$}   
   \dimen0=\wd0                                     
   \setbox1=\hbox{/} \dimen1=\wd1                   
   \ifdim\dimen0>\dimen1                            
      \rlap{\hbox to \dimen0{\hfil/\hfil}}          
      #1                                            
   \else                                            
      \rlap{\hbox to \dimen1{\hfil$#1$\hfil}}       
      /                                             
   \fi}                                             %
\def\ptmiss{\slashchar{p}_{T}}
\def\etmiss{\slashchar{E}_{T}}
\def\stacksymbols #1#2#3#4{\def\theguybelow{#2}
    \def\vp{\lower#3pt}
    \def\sp{\baselineskip0pt\lineskip#4pt}
    \mathrel{\mathpalette\intermediary#1}}

\def\intermediary#1#2{\vp\vbox{\sp
     \everycr={}\tabskip0pt
     \halign{$\mathsurround0pt#1\hfil##\hfil$\crcr#2\crcr
              \theguybelow\crcr}}}

\def\comment#1{}
\def\to{\rightarrow}

\def\u1x{U(1)_X}
\def\vew{v_{EW}}
\newcommand{\nc}{\newcommand}
\nc{\LL}{L} \nc{\vv}{\tilde{v}} \nc{\ccdot}{\!\cdot\!}
\nc{\gsm}{G_{SM}}
\nc{\vfive}{\mathbf{5}\oplus\mathbf{\overline{5}}}
\nc{\vten}{\mathbf{10}\oplus\mathbf{\overline{10}}}
\nc{\zhol}{Z^{\rm hol}}

\begin{document}

\setlength{\baselineskip}{0.2in}



\begin{titlepage}
\noindent
\begin{flushright}
{\small CERN-PH-TH-047} \\
{\small MCTP-09-11}  \\
\end{flushright}
\vspace{-.5cm}

\begin{center}
  \begin{Large}
    \begin{bf}
Dark matter and Higgs boson collider implications \\
of fermions in an abelian-gauged hidden sector\vspace{0.2cm}\\
     \end{bf}
  \end{Large}
\end{center}
\vspace{0.2cm}
\begin{center}
\begin{large}
Shrihari Gopalakrishna$^{a}$, Seung J. Lee$^b$, James D. Wells$^{c}$\\
\end{large}
  \vspace{0.3cm}
  \begin{it}

$^a$~Physics Department, Brookhaven National Laboratory, Upton, NY 11973 
\vspace{0.2cm}\\
$^b$~Dept of Particle Physics, Weizmann Institute of Science, Rehovot 76100 Israel
\vspace{0.2cm}\\
$^c$~CERN Theoretical Physics (PH-TH), CH-1211 Geneva 23, Switzerland, and \\
MCTP, University of Michigan, Ann Arbor, MI 48109
\vspace{0.2cm}\\
 \vspace{0.1cm}
\end{it}

\end{center}


\begin{abstract}

We add fermions to an abelian-gauged hidden sector. We show that the lightest can be the dark matter
with the right thermal relic abundance, and discovery is within reach of upcoming dark matter detectors. We also show that these fermions change Higgs boson phenomenology at the Large Hadron Collider~(LHC), and in particular could induce a large invisible width to the lightest Higgs boson state. Such an invisibly decaying Higgs boson can be discovered with good significance in the 
vector boson fusion channel at the LHC.

\end{abstract}

\vspace{1.4cm}

\flushleft{\today}

\end{titlepage}

\setcounter{page}{2}




\noindent
{\Large\it Abelian-Gauged Hidden Sector}

We work in the context of the theory motivated and developed in 
Refs.~\cite{Schabinger:2005ei,Strassler:2006im,Kumar:2006gm,Bowen:2007ia,Chang:2007ki,MarchRussell:2008yu,Gopalakrishna:2008dv,Feng:2008mu}.
We start with the theory in Ref.~\cite{Gopalakrishna:2008dv}, 
where the Higgs sector Lagrangian is
\bea
{\cal L}_{\Phi} &=& |D_\mu \Phi_{SM}|^2
+ |D_\mu \Phi_H |^2 
 + m^2_{\Phi_H}|\Phi_H|^2 + m^2_{\Phi_{SM}}|\Phi_{SM}|^2 \nonumber\\
& & - \lambda|\Phi_{SM}|^4 - \rho|\Phi_H|^4 - \kappa
|\Phi_{SM}|^2|\Phi_H|^2 \ , \label{Lphi.EQ}
\eea
so that $U(1)_X$ is broken spontaneously by $\left< \Phi_H \right> = \xi/\sqrt{2}$,
and electroweak symmetry is broken spontaneously as usual by
$\left< \Phi_{SM}\right> = (0,\, v/\sqrt{2})^T$.
The Higgs sector mixing is defined by two-dimensional rotation equations
$\phi_{SM}=c_h h+s_h H$ and $\phi_H=-s_h h+c_h H$,  where $h,H$ are the mass eigenstates.
We will take the mixing angle $s_h$ to be an input parameter.

If there are new fermions charged under the $U(1)_X$ but are singlets under 
the SM gauge group, they could impact phenomenology in interesting ways. In particular,
the lightest charged fermion is stable and can be the dark matter. Furthermore, the Higgs bosons
in the theory are expected to mix and induce Higgs phenomenology significantly different from
the SM. Most especially, as we shall show, the Higgs boson will lose significance in all SM channels,
and in addition can decay dominantly into the hidden sector fermions and thus be invisibly decaying. Other related works involving Hidden-sector dark matter are discussed in 
Refs.~\cite{Kim:2008pp,Feldman:2007wj}.

\medskip\noindent
{\Large\it Adding Dirac Fermions}

We consider a theory with two vector-like pairs ($\psi,\ \psi^c$) and ($\chi,\ \chi^c$)
that carry $U(1)_X$ charges but not any SM gauge quantum numbers.
Since there are no fermions charged under both the SM gauge group and $U(1)_X$, there
are no mixed anomalies. The vector-like nature makes the $U(1)_X$ anomaly cancellation
trivial.
We add the Lagrangian terms (written with Weyl spinors)
\bea
{\cal L} \supset & & {\cal L}_\psi^{CD} + {\cal L}_{\psi^c}^{CD} + {\cal L}_\chi^{CD} + {\cal L}_{\chi^c}^{CD} + \nonumber \\
&-&\left( \lambda_s \Phi_H \psi \chi + \lambda_s^\prime \Phi_H^* \psi^c \chi^c + h.c. \right) + \nonumber \\
&-& \left( M_\psi \psi^c \psi + M_\chi \chi^c \chi + h.c. \right) \ ,
\eea
where the covariant derivative term is
\beq
{\cal L}_\psi^{CD} = \psi^\dagger i \bar\sigma^\mu \partial_\mu \psi + g_X \psi^\dagger \bar\sigma^\mu q_\psi \psi \hat{X}_\mu \ ,
\eeq
and similarly for the other covariant-derivative terms; $q_\psi$ represents the $U(1)_X$ charge of $\psi$. 
We assume that the vector-like masses $M_\psi$ and $M_\chi$ are around the electroweak scale.
$U(1)_X$ invariance requires $q_{\Phi_H} + q_{\psi} + q_{\chi} = 0$.
We additionally require $q_{\Phi_H} \neq 0$ since its VEV breaks $U(1)_X$, which then implies that
$q_\psi \neq - q_\chi$. Other than these restrictions, the charges can be 
chosen freely.

There is an accidental $Z_2$ symmetry under which $\psi,\ \psi^c,\ \chi,\ \chi^c$ are odd,
while $\Phi_H$ and all SM fields are even. This ensures the stability of the lightest $Z_2$ odd fermion,
which we will identify as the dark-matter candidate.

In addition to the vector-like masses, $U(1)_X$ breaking by $\left< \Phi_H \right> = \xi/\sqrt{2}$
implies the Dirac masses
$m_D \equiv \lambda_s \xi/\sqrt{2}$ and $m_D^\prime \equiv \lambda_s^\prime \xi/\sqrt{2}$.
We define the Dirac spinors
\beq
\Psi \equiv \bmat \psi \\ \psi^c \emat \quad ; \qquad \Chi \equiv \bmat \chi \\ \chi^c \emat \ ,
\eeq
with the charge-conjugate of these spinors given by $\Psi^c$ and $\Chi^c$.
The mass terms can be written as
\beq
{\cal L}_{\rm mass} = -\bmat {\bar\Psi_R} & {\bar\Chi^c_R} \emat
\bmat M_\psi & m_D^\prime \\ m_D & M_\chi \emat 
\bmat \Psi_L \\ \Chi_L^c \emat + h.c. 
\label{Lmass.EQ}
\eeq
We go to the mass basis $\{ \Psi,\Chi^c\}_{L,R}\to \{ \Psi_1,\Psi_2\}_{L,R}$ by simple two-dimensional
rotations  characterized by the angles $\theta_{L,R}$: $\Psi_L=c_{\psi_L} \Psi_{1L}+s_{\psi_L}\Psi_{2L}$, etc., where $s_\psi, c_\psi$ denote the sine and cosine of the angle respectively. 
There are thus two mass eigenstates, whose masses $M_1$ and $M_2$ are straightforwardly computable from the couplings in the lagrangian above.

The Higgs-$\Psi$-$\Psi$ interactions can be obtained
by replacing $m_D \to m_D (1+\phi_H/\xi)$ and $m_D^\prime \to m_D^\prime (1+\phi_H/\xi)$
in Eq.~(\ref{Lmass.EQ}).  We 
find the couplings between the Higgs mass eigenstates ($h,H$) to the fermion
mass eigenstates ($\psione, \psitwo$) (Feynman rules)
\bea
\bar\Psi_1 \Psi_1 \{ h, H \} \ &:& \ -\frac{i}{\sqrt{2}} \kappa_{11} \{-s_h, c_h\} \nonumber \\
\bar\Psi_2 \Psi_2 \{ h, H \} \ &:& \ -\frac{i}{\sqrt{2}} \kappa_{22} \{-s_h, c_h\}\nonumber \\
\bar\Psi_1 \Psi_2 \{ h, H \} \ &:& \ -\frac{i}{\sqrt{2}} \left( \kappa_{12} P_L + \kappa_{21} P_R \right) \{-s_h, c_h\} \nonumber \\
\bar\Psi_2 \Psi_1 \{ h, H \} \ &:& \ -\frac{i}{\sqrt{2}} \left( \kappa_{21} P_L + \kappa_{12} P_R \right) \{-s_h, c_h\},
\label{hPsiPsi.EQ}
\eea
where we have defined (with all $\kappa$ assumed real)
\bea
\kappa_{11} & = & -\lambda_s^\prime \cR \sL - \lambda_s \sR \cL   \nonumber\\
 \kappa_{12} & = & \lambda_s^\prime \cR \cL - \lambda_s \sR \sL   \nonumber\\
 \kappa_{21} & = &   -\lambda_s^\prime \sR \sL + \lambda_s \cR \cL   \nonumber \\
\kappa_{22} & = &  \lambda_s^\prime \sR \cL + \lambda_s \cR \sL  
\label{kappadef.EQ}
\eea

An alternative theory could be presented with just one vector-like pair of fermions. 
Using Weyl spinors we can write
${\cal L} = -\lambda_m \Phi_H \psi\psi - \lambda_m^\prime \Phi_H^* \psi^c \psi^c - \mpsi \psi\psi^c + h.c.$
This requires the $U(1)_X$ charge assignment $q_H = -2 q_\psi$ and, of course, 
$q_{\psi^c} = - q_\psi$.   
We define a 4-component spinor $\Psi \equiv (\psi\ \psi^c)$ with its 
complex conjugate $\Psi^c$, and after $U(1)_X$ breaking 
we can diagonalize the mass matrix and write the theory in terms of two 
4-component mass eigenstate Majorana spinors $\Psi^M_1$ and $\Psi^M_2$ 
given as linear combinations of $\Psi$ and $\Psi^c$.
The theory so obtained after $U(1)_X$ breaking is 
${\cal L} \supset  \frac{1}{2} \overline{\Psi^M_i} \left( \slashchar{\partial} - {\mpsi}_i \right) \Psi^M_i - \frac{1}{2} \kappa^M_{ij} \phi_H \overline{\Psi^M_i} \Psi^M_j$ (with $i,j=\{1,2\}$)
where the extra factor of $1/2$ is included as usual in defining $\kappa^M$ for a Majorana fermion
in order to cancel the factor of two from the two Wick contractions 
for a Majorana fermion that arises when computing a matrix element.
The $\overline{\Psi^M} \Psi^M \{h,H\}$ Feynman rules are as given in Eq.~(\ref{hPsiPsi.EQ})
with the $\kappa^M$ definition similar to Eq.~(\ref{kappadef.EQ}) except for the
inclusion of a factor of $1/2$ in the left-hand-side for the above reason.
The phenomenology of this theory is qualitatively similar to the Dirac theory discussed above. 
Although our subsequent discussion will 
mainly be focused on the 
two Dirac fermions case described above,
we will comment later on what things change for the Majorana case.

To complete the description of our  Feynman rules conventions we provide the triple Higgs boson interaction vertices.  Our Lagrangian provides, after $\u1x$ and electroweak symmetry breaking,
${\cal L} \supset -\frac{\kappa\xi}{2} \phi_{SM}^2 \phi_H - 
\frac{\kappa v}{2} \phi_{SM} \phi_H^2 - \lambda v \phi_{SM}^3 - \rho\xi \phi_H^3$
which in the mass basis implies the following relevant Feynman rules:
\beq
hhh \ : \ -\frac{i}{\sqrt{2}} v c_h \kappa_{3\phi} \ , \qquad 
Hhh \ : \ -\frac{i}{\sqrt{2}} \xi s_h \kappa_{H2h} \ , 
\label{k3phiDef.EQ}
\eeq
These serve to define the dimensionless cubic couplings $\kappa_{3\phi}$ and $\kappa_{H2h}$, and 
while we can show these in terms of the fundamental Lagrangian parameters, 
it is sufficient for our purposes to treat them as effective input parameters.

\medskip\noindent
{\Large\it Parameters of the Theory} 

We will explore the cosmological, direct-detection and collider implications
of the theory we have outlined. 
We will restrict ourselves to the lightest (and therefore stable) hidden sector fermion
$\Psi_1$ (denoted as $\psi$ henceforth), although many interesting effects
can occur due to transitions to and from more massive states such as the $\Psi_2$. 
We will take an effective theory approach and note that 
the phenomenology is identical to a large class of theories with a hidden sector fermion $\psi$
interacting via the Higgs in the way we have outlined. 
The relevant parameters are: $\mpsi$, $\kappa_{11}$, $\kappa_{3\phi}$, $s_h$ and $\mh$. 

Partial-wave unitarity imposes upper limits on combinations of the parameters 
$\kappa_{11}$, $s_h$, $\kappa_{3\phi}$. For instance, the total cross-section
$\sigma(\psi\psi\to YY)$, where $YY$ generically denotes a pair of final state
particles, is bounded by unitarity~\cite{Yao:2006px} as
$\sigma_{\ell} < 16\pi (2\ell+1)/s$, 
where $\sigma_{\ell}$ denotes the cross-section in the $\ell^{\rm th}$ partial-wave,
and $s$ is the usual Mandelstam variable.
To correctly obtain the bound one needs to impose the bound on the partial waves,
but we present below a sufficient condition using the total cross-section to show that
the parameter ranges we will consider in this work are safe with respect to
unitarity constraints.  
For non-relativistic $\psi$, assuming that the unitarity bound is
saturated by $\sigma(\psi\psi\to\psi\psi)$ we find the bound 
$\kappa_{11} s_h \lesssim 2.5$, with the 
other final-states giving weaker bounds for $\mpsi < 160$~GeV.
For $\mpsi > 160$~GeV, $\sigma(\psi\psi\to W^+ W^-)$ gives the strongest bound
$\kappa_{11} s_h c_h \lesssim 1$. 
Also, when kinematically allowed, $\sigma(\psi\psi\to h h)$ gives an additional constraint
that is rather weak $\kappa_{11}\kappa_{3\phi} s_h c_h \lesssim 10^4$.
We further note that the behavior of the cross-sections
is such that the bound only gets weaker as the CM energy becomes large
compared to $\mpsi$ and $\mh$. 

\medskip\noindent
{\Large\it Relic density}

$\psi\psi$ annihilations into
the $W^+W^-$, $Z Z$, $h h$, $t\bar t$ final states will be important
if they are kinematically accessible, 
and if not, the dominant channel is into $b\bar b$. 
We compute the annihilation cross-section  
in the mass basis including $s$, $t$ and $u$-channel graphs.

The $\psi$ are non-relativistic during freeze-out and
the annihilation cross-sections can be written in the non-relativistic limit~\cite{Wells:1994qy}
to leading order in $|{\bf p}_\psi|$, the 3-momentum magnitude of the incoming $\psi$, 
with $|{\bf p}_\psi|^2/\mpsi^2 = v_{rel}^2/4 + O(v_{rel}^4)$, 
$s = s_0 (1+ v_{rel}^2/4) + O(v_{rel}^4)$, where $s_0 \equiv 4 \mpsi^2$, 
and $v_{\rm rel}$ is the relative velocity between the colliding $\psi$. 
We can write 
$\sigma v_{rel}  = a + b~ v_{rel}^2 + O(v_{rel}^4)$,  
defining $a$ and $b$ as used commonly in the literature. 
The thermally averaged cross-section is then given by~\cite{thermal average}: 
$\left<\sigma v\right>(x_f) \approx a + (6 b-9a)/x_f$ during freeze-out, where $x_f\equiv M_\psi/T_f$ is the unitless measure of the freeze-out temperature $T_f$.

Once the thermally averaged cross-section $\left<\sigma v\right>$ is obtained,
we can compute the ratio of the 
present relic density to the critical energy density (for summaries, see 
for example Refs.~\cite{Bertone:2004pz,Servant:2002aq,Gopalakrishna:2006kr}),
which is given by
\beq
\Omega_0 h^2 =  x_f  ~ \frac{10^{-29}\, {\rm eV}^{-2}}{\langle \sigma v\rangle(x_f)} \ ,
\label{Om0sig.EQ}
\eeq
where $h^2 \approx 0.5$,
and we can take $x_f \approx 25$ to a good approximation since it depends 
only mildly (logarithmically) on the parameters. To obtain
the observed dark matter relic density we need 
$\left<\sigma v \right> \sim 1.5 \times 10^{-9}~{\rm GeV^{-2}}$.

Next, we present analytical formulas for the self-annihilation cross-section into
the dominant channels.
In the annihilation cross-sections below, we will omit showing the decay widths of the 
particles in the propagators, 
and also not show the heavy Higgs contribution since we take $\mH \gg \mpsi$,
although we will include it in the numerical analysis to be presented.

In the center-of-mass (CM) frame, the annihilation cross-section of a pair of 
Dirac $\psi$ into SM $f \bar f$ is given by
\beq
\sigma\left( \psi \bar\psi \to f \bar f\right) \approx
\frac{N_c \kappa_{11}^2 \lambda_f^2 s_h^2 c_h^2}{8 \pi v_{\rm rel}} 
\frac{|{\bf p}_\psi|^2}{\left(s - m_h^2 \right)^2} 
\left( 1-\frac{4 m_f^2}{s} \right)^{3/2} \ , 
\label{sigSAff.EQ}
\eeq
where $N_c=3$ for a fermion in the fundamental of $SU(3)_c$, 
and $\lambda_f/\sqrt{2}$ is the $\phi_{SM} f \bar f$ Yukawa coupling.
Eq.~(\ref{sigSAff.EQ}) is valid for both $b\bar b$ and $t\bar t$ final states.
For $\mpsi > \mw$, the $W^+W^-$ channel is accessible, and its cross-section is given by
\bea
\sigma(\psi \bar\psi \to W^+ W^-) \approx \frac{\kappa_{11}^2 g^4 s_h^2 c_h^2}{8 \pi v_{\rm rel}} 
\frac{\vew^2}{s} 
\frac{|{\bf p}_\psi|^2}{\left(s - m_h^2 \right)^2 } 
\sqrt{1-\frac{4\mw^2}{s}} \left[\frac{1}{2}+\frac{(s/2-\mw^2)^2}{4\mw^4}\right]  \ ,
\label{sigSAww.EQ}
\eea
where $\vew = 246$~GeV. 
For $\mpsi > \mz$, the $ZZ$ final state will be accessible also, and 
the annihilation cross-section is similar to Eq.~(\ref{sigSAww.EQ})
but with an extra factor of $1/2c_W^4$, and with $\mw\to\mz$. 
For $\mpsi > \mh$ the $h h$ final state will be open, 
and including the $s$, $t$ and $u$-channel graphs, we find the
cross-section
\bea
\sigma(\psi \bar\psi \to h h) & \approx & \frac{\kappa_{11}^2 s_h^2}{8 \pi v_{\rm rel}}
\frac{|{\bf p}_\psi|^2}{\mpsi^2}  
\sqrt{1-\frac{4m_h^2}{s}} 
\left\{
\frac{\kappa_{3\phi}^2 c_h^2 \vew^2}{16\left(s-m_h^2 \right)^2} 
- \frac{\kappa_{3\phi} \kappa_{11} c_h s_h \vew \mpsi}{2(s-m_h^2)(t_0 - \mpsi^2)}
\right. \nonumber \\
& & \left. \cdot \left[ 1-\frac{t_0}{3\left(t_0-\mpsi^2 \right)}  \right] +
\frac{\kappa_{11}^2 s_h^2 \mpsi^2}{(t_0-\mpsi^2)^2}
\left[ 1-\frac{t_0}{12\left(t_0-\mpsi^2 \right)}  \right]
\right\}
\label{sigSAhh.EQ}
\eea
with the dimensionless Higgs cubic coupling $\kappa_{3\phi}$ as given in 
Eq.~(\ref{k3phiDef.EQ}). 
We have defined $t_0\equiv -|{\bf k_h}|^2$ to be the Mandelstam variable $t$ in the 
${\bf p_\psi}\to 0$ limit with ${\bf k_h}$ being the three momentum of the outgoing Higgs boson.  

The reason that all cross-sections above are proportional to $|{\bf p}_\psi|$ follows 
from the $CP$ transformation property of the initial, intermediate and final states,
and angular-momentum conservation.  
Under a CP transformation, the two-fermion initial state transforms as
$(-1)^{L+S} \times (-1)^{L+1} = (-1)^{S+1}$, shown split up into the
$C$ and $P$ transformation factors respectively, 
where $L$ is the total orbital angular momentum and $S$ the total spin. 
For the $s$-channel graphs involving the intermediate CP-even scalar Higgs boson $h$,
the above CP transfromation property and angular momentum conservation 
imply that the initial state has to be in $L=1$ (p-wave) and $S=1$ configuration. 
For the $t$-channel graph into the $hh$ final state 
the CP transformation property, angular momentum conservation and the
requirement that the two-boson final state be symmetric under interchange
implies that the two-fermion initial state has to again be in the $L=1$ ($p$-wave) and $S=1$
configuration. 
In either case, since the 2-fermion initial state has to be in a $p$-wave configuration 
the matrix-element goes to zero as $|{\bf p}_\psi|\rightarrow 0$.
Therefore, the coefficient $a$ is zero
since there is no velocity independent piece in the annihilation cross-section.
 
\begin{figure}[ttt]
\begin{center}
\includegraphics[angle=0,width=0.42\textwidth]{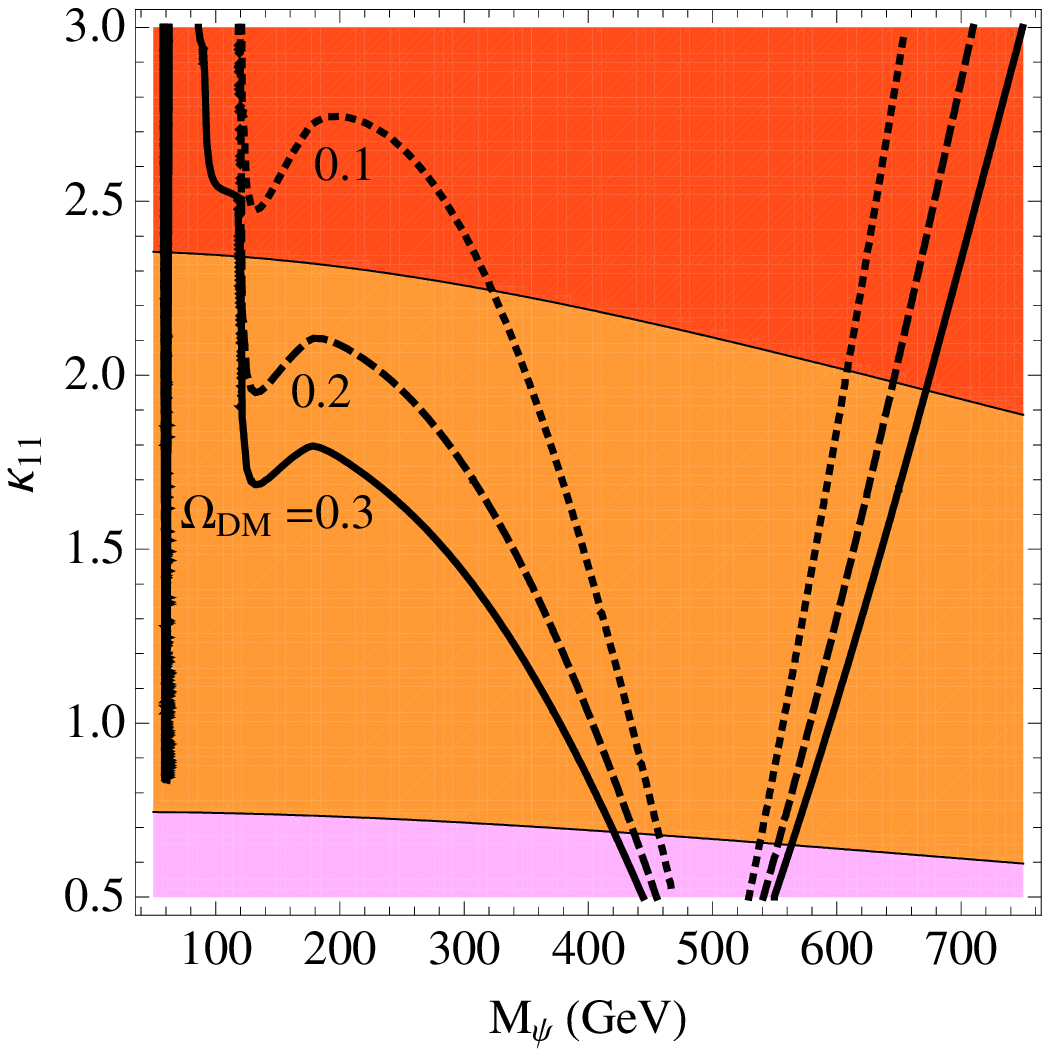}
\hspace*{0.25cm}
\includegraphics[angle=0,width=0.42\textwidth]{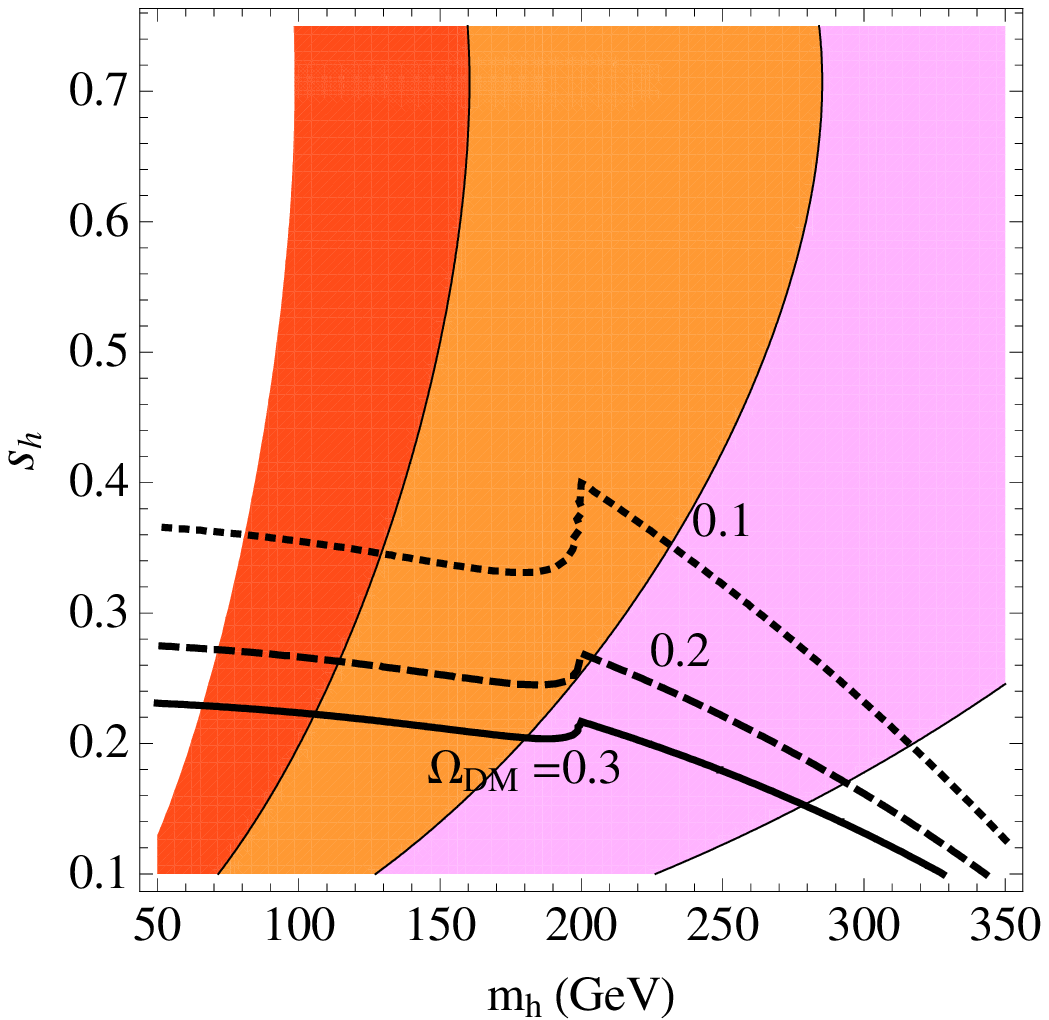}
\caption{
Contours of ${\Omega_{dm}}_0 = 0.1,~0.2,~0.3$ (dot, dash, solid) in the 
$\mpsi$--$\kappa_{11}$ and $\mh$--$s_h$
planes.
The parameters not varied in the plots are fixed at
$\mpsi = 200~$GeV, $\mh = 120~$GeV, $s_h=0.25$, $\kappa_{11}=2$,  
$\kappa_{3\phi} = 1$, $\mH=1~$TeV, $\kappa_{H2h}=1$ and $\xi=1~$TeV.
The direct-detection $\psi-N$ cross-section are shown as shaded regions:
$\sigma \gtrsim 10^{-43}~{\rm cm}^2$ (dark-shade)
is already excluded by experiments. 
$\sigma \gtrsim 10^{-44}~{\rm cm}^2$ (medium-shade),
and $\sigma \gtrsim 10^{-45}~{\rm cm}^2$ (light-shade), the latter two
will be probed in upcoming experiments.  
\label{om0dep.FIG}}
\end{center}
\end{figure}

We show in Fig.~\ref{om0dep.FIG} the (0.1,~0.2,~0.3) contours of ${\Omega_{dm}}_0$ 
in the $\mpsi$--$\kappa_{11}$ and $\mh$--$s_h$ 
planes,
with the parameters not varied in the plots fixed at 
$\mpsi = 200~$GeV, $\mh = 120~$GeV, $s_h=0.25$, $\kappa_{11}=2.0$,  
$\kappa_{3\phi} = 1$, $\mH=1~$TeV, $\kappa_{H2h}=1$ and $\xi=1~$TeV.
This bench-mark point results in $\Omega_{dm} \approx 0.2$. 
The present experimental data on the
dark matter relic density is $\Omega_{dm} = 0.222 \pm 0.02$, 
inferred from the following data~\cite{Yao:2006px}:
total matter density $\Omega_m h^2 = 0.132 \pm 0.004$,
baryonic matter density $\Omega_b h^2 = 0.0219 \pm 0.0007$,
and $h = 0.704\pm 0.016$. 
We see that there exists regions of parameter space that are consistent with 
the present experimental observations.
The region $\Omega_0 < 0.2$ is still allowed but without
the $\psi$ being all of the dark matter, while the 
region $\Omega_0 > 0.3$ is excluded since the $\psi$ 
would overclose the universe. The contours funnel-down at $M_\psi\simeq 500\gev$ due to a resonant annihilation through the heavy Higgs boson, which is taken here to be $1\tev$.
In the region $\mh > 2 \mpsi$, the $h\to \psi\psi$ decay is allowed,
implying an invisibly decaying Higgs at a collider. This connection
will be explored in a later section.

The Majorana case is qualitatively similar to the Dirac case above.
In relating to the number density of the dark matter relic,
there is only one species for the Majorana case 
(rather than two, particle and anti-particle, for the Dirac case), implying a relative 
factor of $1/2$ in the $\left< \sigma v \right>$ for the Majorana case.  
(As we have already noted, the coupling for the Majorana case is defined with a relative
factor of $1/2$ to cancel the factor of two from the two Wick contractions.)
The exact values of the Lagrangian parameters that will result in the 
correct relic density will therefore be different, but will result in 
qualitatively similar plots. We will therefore not repeat the plots for this case.

\medskip\noindent
{\Large \it Direct Detection of Dark Matter} 

Many experiments are underway currently to directly detect dark matter, and still
more are proposed to improve the sensitivity. 
In order to ascertain the prospects of directly observing $\psi$ in the $\u1x$ framework we
are considering, we compute the elastic $\psi$-nucleon cross-section
due to the $t$-channel exchange of the Higgs boson. 
The typical dark-matter velocity in our galaxy is about $270$~km/s~\cite{Bertone:2004pz},
which implies $v_\psi \sim 10^{-3}$, and the $\psi$ is quite non-relativistic.
In the CM frame, in the non-relativistic limit, we find the elastic cross-section  
\beq
\sigma\left(\psi N \to \psi N \right) \approx
\frac{\kappa_{11}^2 s_h^2 c_h^2 \lambda_N^2}{8\pi v_{rel}} 
\frac{(|{\bf p}_\psi|^2 + m_N^2)}{m_h^4} \ ,
\eeq
where $|{\bf p}_\psi| \approx \mpsi v_\psi$, $m_N \approx 1$~GeV is the nucleon mass,
$\lambda_N/\sqrt{2}$ is the effective $h\bar N N$ coupling, and we have ignored
the Mandelstam variable $t$ in comparison to $m_h^2$ in the Higgs propagator
which give a contribution of order $|{\bf p}_\psi|^2 m_N^2/m_h^2$. 
We take $\lambda_N\approx 2\times 10^{-3}$~\cite{Shifman:1978zn,Bertone:2004pz}
which includes the Higgs tree-level coupling to light quarks ($u,d,s$), 
and the heavy-quark-loop two-gluon couplings. 

To illustrate, for $\kappa_{11}=2.0$, $s_h=0.25$, $\mpsi=200$~GeV, $\mh=120$~GeV, 
we find $\sigma \approx 1.9\times 10^{-16}~{\rm GeV}^{-2} = 7\times 10^{-44}~{\rm cm}^2$.
This is very interesting as the presently ongoing experiments~\cite{DirDetExptCite} 
are probing this range of cross-sections. 
With all other parameters fixed as above, as $\mh$ is increased to $350$~GeV, 
the direct-detection cross-section falls smoothly to about $10^{-45}~{\rm cm}^2$. 
In Fig.~\ref{om0dep.FIG}
we show the direct detection cross-section as shaded regions;
from the compilation in Ref.~\cite{DirDetExptCite}, 
the dark-shaded region ($\sigma \gtrsim 10^{-43}~{\rm cm}^2$) is excluded by 
present bounds from direct detection searches~\cite{CDMS, Xenon}, 
while the medium-shaded ($\sigma \gtrsim 10^{-44}~{\rm cm}^2$) and the 
light-shaded ($\sigma \gtrsim 10^{-45}~{\rm cm}^2$)
regions will be probed by upcoming experiments, 
such as Super-CDMS and Xenon 1-ton~\cite{future}.
We have defined our model into the package MicrOMEGAs~\cite{Belanger:2006is} 
and checked that our analytical results 
agree with the full numerical treatment reasonably well.

\medskip\noindent{\Large\it Higgs Boson Decays}

In addition to the usual SM decay modes,
if $\mpsi < \mh/2$, the decay $h \to \psi \bar\psi$ is kinematically allowed,
leading to an invisible decay mode for the Higgs boson.
Here, we explore how the Higgs decay is affected if this is the case. One should note that if 
kinematically allowed, the Higgs can decay into a pair of $\u1x$ gauge fields 
($X_\mu$), which was the subject of Ref.~\cite{Gopalakrishna:2008dv}.
We will not include the $h\to XX$ decay mode explicitly in our analysis here,
but in regions of parameter space where this is present, its effect would be to decrease
all branching ratios discussed in this work.
For simplicity, we will consider only on-shell 2-body decays and
do not include virtual 3-body modes. Also, we will use the narrow-width
approximation in all decay chains.

The Higgs decay width is easily computed using the coupling in Eq.~(\ref{hPsiPsi.EQ}). 
For a Dirac $\psi$, the invisible decay width of the Higgs boson is given by
\beq
\Gamma{\left(h\to \psi \bar\psi \right)} = 
\frac{\kappa_{11}^2 s_h^2}{16\pi} \mh \left(1-\frac{4\mpsi^2}{m_h^2}\right)^{3/2} \ .
\eeq
The partial decay width to a SM fermion pair $f\bar f$ is given by
\beq
\Gamma\left(h\to f \bar f \right) = \frac{N_c g^2 \mf^2 c_h^2}{32\pi \mw^2} \mh 
\left(1-\frac{4\mf^2}{m_h^2}\right) \ ,
\eeq
where $N_c=3$ is for a fermion in the fundamental of $SU(3)_c$. 
The Higgs partial width to a SM gauge boson pair is given by
\bea
\Gamma\left(h\to W^+ W^- \right) &=& \frac{g^2 c_h^2}{64\pi \mw^2} m_h^3 
\sqrt{1-\frac{4\mw^2}{m_h^2}}\left(1-\frac{4\mw^2}{m_h^2} + \frac{12\mw^4}{m_h^4} \right) \ , \\
\Gamma\left(h\to Z Z \right) &=& \frac{g^2 c_h^2}{128\pi \mw^2} m_h^3 
\sqrt{1-\frac{4\mz^2}{m_h^2}}\left(1-\frac{4\mz^2}{m_h^2} + \frac{12\mz^4}{m_h^4} \right) \ .
\eea

\begin{figure}[ttt]
\begin{center}
\includegraphics[angle=0,width=0.45\textwidth]{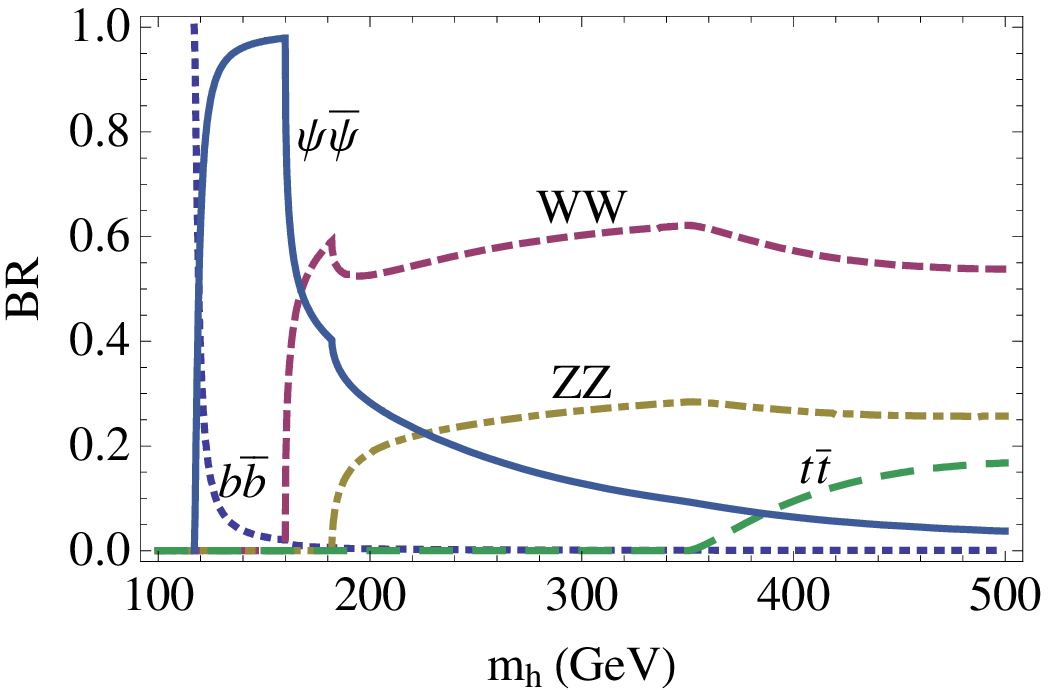}
\hspace*{0.25cm}
\includegraphics[angle=0,width=0.47\textwidth]{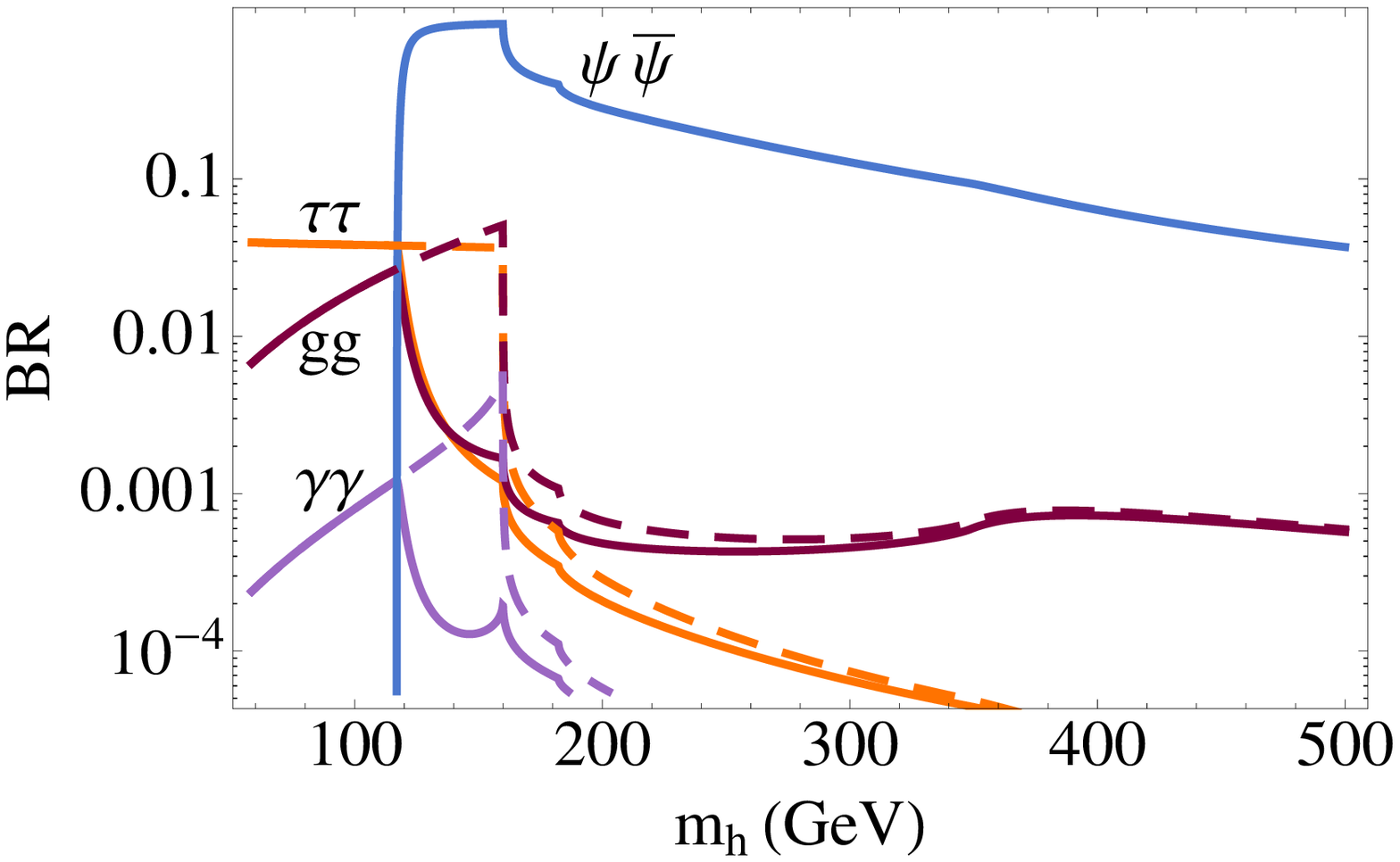}
\caption{
The left plot shows the Higgs BR into invisible, $bb$, $WW$, $ZZ$ and $tt$ 
(thick-solid, dotted, dashed, dash-dot and thin), 
as a function of the Higgs mass,
and in the right plot, the solid curves show the light Higgs BR in the $\u1x$ model, 
with the dashed curves showing the SM Higgs BR's.
The other parameters are fixed at $\mpsi = 58.5$~GeV, $s_h=0.25$, $\kappa_{11}=2.0$,
$\kappa_{3\phi}=1.0$ and $\mH=1$~TeV. 
\label{hGamBR.FIG}}
\end{center}
\end{figure}

As mentioned, the $h\to \psi\psi$ offers a new decay mode in the $\u1x$ model,
and in Fig.~\ref{hGamBR.FIG}~(left) we show the Higgs BR, 
and a comparison to a few SM modes (right),
with the other parameters fixed at 
$\mpsi=58.5$~GeV, $s_h=0.25$, $\kappa_{11}=2.0$,
$\kappa_{3\phi}=1.0$ and $\mH=1$~TeV. 
These parameter values result in the correct relic-density 
for $m_h=120~$GeV. 
In this light Higgs and small $\mpsi$ region where the $b\bar b$ channel is
the dominant final-state, an acceptable relic-density is obtained only when
$\mh \approx 2 \mpsi$, i.e. when the Higgs boson pole enhances
the cross-section which otherwise is generically too small 
due to the small $b$ Yukawa coupling. Thus, for the $\mh$ range shown in the figure
and for the parameter values shown above, 
the correct relic-density is obtained only for $\mh\approx 120~$GeV. 

For a relatively light Higgs boson ($\mh < 2 \mw$) and with $\mpsi\lesssim \mh/2$
the invisible BR dominates in the $\u1x$ scenario. 
For $\mpsi \lesssim 60$~GeV, we find $\Gamma_h(\mh=120~{\rm GeV}) \approx 5\times 10^{-3}$~GeV
and $\Gamma_h(\mh=200~{\rm GeV}) \approx 2$~GeV. 
For smallish $s_h$, the invisible BR is not as large for a heavier Higgs boson since the SM Higgs 
boson already has a sizable width due to $h\to W^+ W^-$, etc. 
When the invisible BR is large, as we will show in greater detail in the following, the 
SM BRs, for example, into $b\bar b$ and the $\tau^+\tau^-$ are suppressed,
implying that the standard search channels at the LHC will 
have a reduced significance. We will however show that the invisible mode
holds promise for the discovery of the Higgs in this scenario. 

\begin{figure}[ttt]
\begin{center}
\includegraphics[angle=0,width=0.45\textwidth]{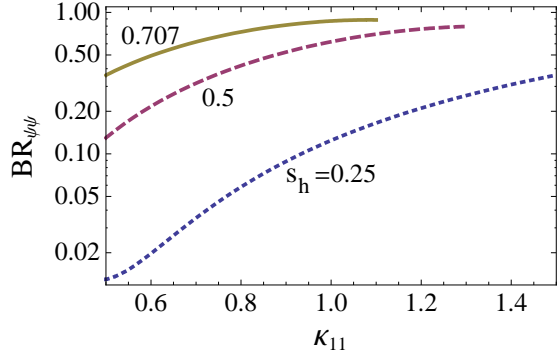}
\caption{The \brinv as a function of $\kappa_{11}$ for $\mh=120$~GeV
for $s_h=0.25,0.5,0.707$ (dotted, dashed, solid) with $\mpsi$ 
adjusted to give the correct dark matter relic density ($\Omega_0$).
The other parameters are fixed at $\kappa_{3\phi}=1.0$ and $\mH=1$~TeV.
\label{hBRinv.FIG}}
\end{center}
\end{figure}

The parts of the parameter space that yield the correct dark matter relic density
have been discussed earlier (see Fig.~\ref{om0dep.FIG}). 
We impose the requirement that the relic density should 
be in the experimentally measured range by scanning over 
$\mpsi \sim 60$~GeV, and show in 
Fig.~\ref{hBRinv.FIG} the corresponding \brinv as a function of $\kappa_{11}$ (see Eq.~(\ref{hPsiPsi.EQ})), 
with $\kappa_{3\phi}=1.0$ and $\mH=1$~TeV held fixed.
We see that a significant \brinv is possible while giving the
required $\Omega_0$ and being consistent with present direct-detection limits,
with the general trend
of increasing \brinv for increasing $\kappa_{11}$ or $s_h$. 
Here we have shown only the points that satisfy the direct-detection 
cross-section $\sigma < 10^{-43}~{\rm cm}^2$,
to be consistent with current experimental results~\cite{DirDetExptCite}.
For a larger Higgs mass we find qualitatively similar invisible BR with
larger values of $\kappa_{11}$ preferred. 

For the Majorana case, 
since the two
final state particles are identical, the phase-space integration results
in a factor of $1/2$ compared to the Dirac case.
Therefore the decay rate for the Majorana case is $1/2$ of the Dirac case. 
As we have already commented, note that the coupling in the Majorana case 
is defined with a factor of $1/2$ compared to the Dirac case. 
Again, we will not repeat the plots for the
Majorana case since they are qualitatively similar to the Dirac case
presented previously.

\medskip\noindent{\Large\it LHC Higgs Boson Phenomenology}

In order to see how the suppression of the SM modes will affect the 
significance in the standard search channels, 
we estimate the ratio of the discovery significance of the
light Higgs in the $gg\rightarrow h \rightarrow \gamma\gamma$, 
$gg\rightarrow h \rightarrow ZZ\rightarrow 4\ell$ and
$gg\rightarrow h \rightarrow WW\rightarrow 2\ell2\nu$
channels to those
of a SM Higgs boson with the same mass. 
An approximate formula for the
ratio of the Higgs discovery significance in the $\u1x$ model compared to that in the SM 
for the same mass in the
$gg\rightarrow h\rightarrow XX$ channel can be defined as
\begin{eqnarray}
R_S^{XX}&\equiv&\frac{S(h)}{S(h_{SM})}=\frac{\Gamma(h\rightarrow g g)\,B(h\rightarrow XX)}{\Gamma(h_{SM}\rightarrow gg)\,B(h_{SM}\rightarrow XX)}\cdot F_{XX}(\Gamma)
\end{eqnarray}
where 
\begin{eqnarray}
F_{XX}(\Gamma)=\sqrt{\frac{\rm{max}(\Gamma_{tot}(h_{SM}),\Delta M_{XX})}{\rm{max}(\Gamma_{tot}(h),\Delta M_{XX})}} \, ,
\label{sigXSM.EQ}
\end{eqnarray}
 if the final state is a resonance (i.e., $\gamma\gamma$ or $4l$) and 
$F_{XX}(\Gamma)=1$ otherwise.
Although as the Higgs boson width gets smaller the signal to background ratio increases, 
the finite detector resolution of the 
invariant mass ($\Delta M_{XX}$) limits this, 
which is accounted for by $F_{XX}$.
%
\begin{figure}[ttt]
\begin{center}
\includegraphics[angle=0,width=0.45\textwidth]{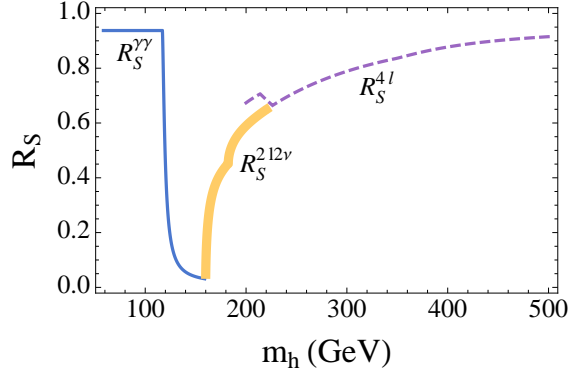}
\caption{The ratio of the LHC Higgs discovery significance in the $\u1x$ model
to that in the SM for $\mpsi = 58.5$~GeV, $\kappa_{11}=2.0$, $s_h=0.25$,
$\kappa_{3\phi}=1$ and $\mH=1$~TeV,
in the $h\to\gamma\gamma$, 
$h\to ZZ\to 4\ell$ and $h\to WW\to 2\ell 2\nu$ channels. 
\label{LHCdetri.FIG}}
\end{center}
\end{figure}

Fig.~\ref{LHCdetri.FIG} shows $R^{\gamma\gamma , 4\ell , 2\ell 2\nu }_{S}$. 
The $h\to\gamma\gamma$ channel
is the primary discovery channel for $\mh\lesssim 150$~GeV in the SM, and
we see that when the invisible Higgs BR is large, the significance in this channel 
in the $\u1x$ model deteriorates as anticipated.   
In the SM, the $4\ell$ channel is the most important discovery channel for 
$\mh\gtrsim 200$~GeV, and we find from $R^{4\ell}_{S}$ in our model 
that for $s_h=0.25$ this channel is still viable, 
but for larger mixing angle, $s_h=0.5$, that it is not  
until $m_h\approx 350$ GeV.
From $R^{2l2\nu}_{S}$ we see that 
for the Higgs mass between $160-200$~GeV the $h\rightarrow WW\rightarrow 2l2\nu$ channel,
which is the most important channel in the SM, loses 
its efficiency. 
Since the Higgs decay channels into SM modes diminish in significance, we turn next to 
the prospects of a new channel -- the invisible decay mode -- as a means of discovering the Higgs.

\medskip\noindent{\Large\it Invisible Decays of the Higgs Boson}

To detect an invisibly decaying Higgs boson, we have to look at associated production 
in order to trigger on the event. Here we consider the $j j h$ channel
(vector boson fusion)~\cite{Eboli:2000ze,Cavalli:2002vs}, and
$Z h$ associated production~\cite{Davoudiasl:2004aj,Zhu:2005hv,ZHinvTheoOth,ZHinvExpt}. 
The $t\bar t h$ channel~\cite{ttHinv} is also a possibility but we will not
discuss this here. 
An invisibly decaying Higgs has also been discussed in other contexts in
Refs.~\cite{Hundi:2006rh, Cao:2007rm}.

{\it $j$\,$j$\,$h$ channel:}
The Higgs boson can be produced via vector boson fusion at the LHC, followed by 
the invisible decay of the Higgs boson. The signature for this mode is two forward tagging jets
plus missing energy, i.e., $j \ j + \etmiss$.
This channel has been analyzed in Ref.~\cite{Eboli:2000ze}, which we use to
obtain significances in the $\u1x$ model by multiplying the signal cross-section
given there by \brinv\-$c_h^2$.
The backgrounds included there are QCD and EW $Z j j$ and $W j j$. 
In Table~\ref{hinv-wbf.TAB} we show the signal ($\sigma_S$) and
background ($\sigma_B$) cross-sections, after the cuts
\bea
& & p_T^j > 40 \ ,\ |\eta_j| < 5.0 \ ,\ |\eta_{j_1} - \eta_{j_2}| > 4.4 \ ,\ \eta_{j_1}\cdot \eta_{j_2} < 0 \ ,  \nonumber \\
& & \ptmiss > 100~{\rm GeV} \ ,\ M_{jj} > 1200~{\rm GeV} \ ,\ \phi_{jj} < 1 \ .
\label{cutswbf.EQ}
\eea
The luminosity required for $5\,\sigma$ statistical significance (${\cal L}_{5\,\sigma}$) in the
$\u1x$ model scales as $1/(BR_{inv}^2 c_h^4)$ which we have  
factored out in the last column. 
\begin{table}
\begin{center}
\caption{
The $p p \to j j h \to j j \etmiss$ channel vector boson fusion signal 
and background cross-sections from Ref.~\cite{Eboli:2000ze}, 
and the luminosity required for $5\,\sigma$ significance.
These are after the cuts shown in Eq.~(\ref{cutswbf.EQ}).
\label{hinv-wbf.TAB}}
\begin{tabular}{|c|c|c|c|}
\hline 
$\mh$ (GeV)&
$\sigma_{S}$ ($fb$)&
$\sigma_{B}$ ($fb$)&
${\cal L}_{5\,\sigma}$ ($fb^{-1}$)\tabularnewline
\hline
\hline 
$120$&
$97\cdot c_h^2\, BR_{inv}$&
$167$&
$0.44\, / \, (BR_{inv}^{2}\, c_{h}^{4})$\tabularnewline
\hline 
$200$&
$77\cdot c_h^2\, BR_{inv}$&
$167$&
$0.7\, / \, (BR_{inv}^{2}\, c_{h}^{4})$\tabularnewline
\hline 
$300$&
$56\cdot c_h^2\, BR_{inv}$&
$167$&
$1.3\, / \, (BR_{inv}^{2}\, c_{h}^{4})$\tabularnewline
\hline
\end{tabular}
\end{center}
\end{table}
For example, for $\mh=120$~GeV, \brinv=0.75 and $s_h=0.5$, we would require a luminosity
of $1.4~{\rm fb}^{-1}$ for $5\,\sigma$ statistical significance. 
Alternatively, with $10~{\rm fb}^{-1}$, 
we can probe \brinv down to about $26$\,\% at $5\,\sigma$.
We thus see that in this channel, the
prospect of discovering an invisibly decaying Higgs boson in the $\u1x$ scenario is excellent. 
The significance remains quite good even for heavier Higgs masses
as can be seen from Table~\ref{hinv-wbf.TAB}. 
Detailed experimental analyses that take into account 
additional QCD backgrounds and 
detector effects are being analyzed~\cite{WBFExpt}. 

{\it $Z$\,$h$ channel:}
In the $Z h$ channel, we focus on the leptonic decay mode of the $Z$, giving the 
signature $\ell^+\ell^-+\etmiss$. 
This has been analyzed in Ref.~\cite{Davoudiasl:2004aj} for 
the $\mh = 120,~140,~160$~GeV cases using the cuts 
\beq
{p_T}_{\ell} > 10 \ ,\ |\eta_\ell| < 2.5 \ ,\ \ptmiss > 100~{\rm GeV} \ ,\ 
|M_{\ell^+\ell^-} - \mz | < 10~{\rm GeV} \ .
\label{cutshll.EQ}
\eeq
We adopt the same cuts given in Eq.~(\ref{cutshll.EQ}) and compute the signal cross-section 
using the Monte Carlo package CalcHEP~\cite{CalcHEP}. 
We use the $ZZ$, $WW$, $ZW$, and $Z+j$
background cross-sections given in Ref.~\cite{Davoudiasl:2004aj}, 
where it is pointed out that 
with the large $\ptmiss > 100~{\rm GeV}$ cut,
the $Z+j$ background~\cite{Frederiksen:1994me} is adequately small.
We show the signal ($\sigma_S$) and background ($\sigma_B$) cross-sections
after cuts in Table~\ref{hinv-hz2hll.TAB}. 
\begin{table}
\begin{center}
\caption{
The $p p \to Z h \to \ell^+\ell^- \etmiss$ channel signal and background 
cross-sections and the luminosity required for $5\,\sigma$ significance.
The background cross-section is from Ref.~\cite{Davoudiasl:2004aj}. 
These are after the cuts shown in Eq.~(\ref{cutshll.EQ}).
\label{hinv-hz2hll.TAB}}
\begin{tabular}{|c|c|c|c|}
\hline 
$\mh$ (GeV)&
$\sigma_{S}$ ($fb$)&
$\sigma_{B}$ ($fb$)&
${\cal L}_{5\,\sigma}$ ($fb^{-1}$)\tabularnewline
\hline
\hline 
$120$&
$9\cdot c_h^2\, BR_{inv}$&
$26.3$&
$8\, / \, (BR_{inv}^{2}\, c_{h}^{4})$\tabularnewline
\hline 
$200$&
$3.4\cdot c_h^2\, BR_{inv}$&
$26.3$&
$58\, / \, (BR_{inv}^{2}\, c_{h}^{4})$\tabularnewline
\hline 
$300$&
$1.1\cdot c_h^2\, BR_{inv}$&
$26.3$&
$543\, / \, (BR_{inv}^{2}\, c_{h}^{4})$\tabularnewline
\hline
\end{tabular}
\end{center}
\end{table}
We have checked that our signal cross-section for $\mh=120$~GeV agrees with 
that in Ref.~\cite{Davoudiasl:2004aj}. 
From Table~\ref{hinv-hz2hll.TAB}, we see for example, 
for $\mh=120$~GeV, \brinv=0.75 and $s_h=0.5$, we would require a luminosity
of $25~{\rm fb}^{-1}$ for $5\,\sigma$ statistical significance.
Alternatively, with $100~{\rm fb}^{-1}$, 
we can probe \brinv down to about $38$\,\% at $5\,\sigma$.
As Table~\ref{hinv-hz2hll.TAB} shows, the luminosity required becomes rather large
for heavier Higgs masses. 

In conclusion, we have shown that fermions in an abelian-gauged hidden sector 
can be the dark-matter observed cosmologically since they are stable due to
an accidental $Z_2$ symmetry, and the experimentally observed 
relic-density is obtained for natural values of the Lagrangian parameters. 
The prospects for directly detecting these fermions in upcoming experiments 
are excellent. We showed that these fermions can potentially be discovered at the LHC
by looking for invisibly decaying Higgs bosons in the vector boson fusion channel.

\medskip{\it Acknowledgments:} We thank S.~Dawson, B.~Kilgore, F.~Paige, A.~Rajaraman, 
C.~Sturm, T.~Tait and C.~Wagner for 
valuable discussions. We also thank A.~Pukhov for help with micrOMEGAs. 
SG is supported in part by the DOE grant DE-AC02-98CH10886 (BNL).
We thank KITP, Santa Barabara, for hospitality during the 
``Physics of the LHC" workshop where part of this work was carried out.

\end{document}